\documentclass[12pt]{iopart}
\usepackage[colorlinks,linkcolor=blue,citecolor=blue,urlcolor=blue]{hyperref}
\usepackage{graphicx}
\usepackage{array}
\usepackage{caption}
\usepackage{subcaption}
\usepackage{float}
\usepackage{cite}
\usepackage{slashed}
\usepackage{iopams}
\begin{document}

\title{Zeno crossovers in the entanglement speed  of  spin chains with noisy impurities}
\date{\today}

\author{Abhijit P. Chaudhari$^{1,2}$, Shane P. Kelly$^2$, Riccardo J. Valencia-Tortora$^2$, Jamir Marino$^2$}
\address{$^1$Department of Physics, Indian Institute of Technology (Banaras Hindu University), Varanasi - 221005, India}
\address{$^2$Institut f\"ur Physik, Johannes Gutenberg Universit\"at Mainz, D-55099 Mainz, Germany}
\eads{\mailto{apravin.chaudhari.phy16@iitbhu.ac.in}}
\begin{abstract}
We use a noisy signal with finite correlation time to drive a spin (dissipative impurity) in the quantum XY spin chain and calculate the dynamics of entanglement entropy of a bipartition of spins, for a stochastic quantum trajectory. We compute the noise averaged entanglement entropy of a bipartition of spins and observe that its speed of spreading decreases at strong dissipation, as a result of the Zeno effect. We recover the Zeno crossover and show that noise averaged entanglement entropy can be used as a proxy for the heating and Zeno regimes. Upon increasing the correlation time of the noise, the location of the Zeno crossover shifts at stronger dissipation, extending the heating regime. 
\end{abstract}
\submitto{\JSTAT}

\maketitle

\section{Introduction}
\label{sec:intro}

Impurity models  represent a traditional avenue~\cite{affleck_quantum_2009, mahan2013many} for developing intuition in complex many-body problems ranging from  condensed matter physics to solid state and encompassing cold atoms. A recent line of investigation is extensively revisiting instances of prototypical impurity systems in a dissipative setting, with the aim to provide future guidance in the solution of the driven-dissipative quantum many-body problem.
The effect of a localised dissipative potential can be implemented, for instance, by shining an electron beam on an atomic BEC~\cite{Ott3, Ott4}, and it results in a decrease of atom losses at strong dissipation rate. This is a manifestation of a many-body version of the Zeno effect~\cite{misra1977zeno,itano1990quantum,facchi2002quantum,kofman1996quantum, kofman2001zeno,kofman2000acceleration, PhysRevB.98.205136, PhysRevLett.126.170602}, where the usual slowdown of transport at strong measurement is at interplay with many-particle correlation effects~\cite{PhysRevLett.122.040402}. Furthermore, losses provided  by a near-resonant optical tweezer at a quantum point contact,  can realise a dissipative scanning gate microscope  for ultra-cold $^6$Li atoms~\cite{lebrat2019quantized,  corman2019quantized}, offering a resource for transport problems. As in their unitary counterparts, dissipative impurities can also have an intrusive effect and   significantly rearrange the properties of many-body states as in the Anderson impurity problem~\cite{mahan2013many,tonielli2019orthogonality}.
A classification of the phenomenology of dissipative impurities is currently subject of vivid research at the interface of quantum information, atomic physics, quantum optics and solid state~\cite{schiro2019quantum,  tonielli2020ramsey, PhysRevA.103.043304, Biella2021manybodyquantumzeno, sartori2015spin, yoshimura2020non, PhysRevResearch.3.L032013, PhysRevLett.126.120603}; it comprises  driven-dissipative boundary spin problems~\cite{vzunkovivc2010explicit, vznidarivc2011solvable, prosen2015matrix, vznidarivc2015relaxation, berdanier2019universal,Bu_a_2020,PhysRevB.105.054303},  and the realisation of the many-particle Zeno effect in lossy (and noisy) condensed matter systems and cold atoms~\cite{PhysRevB.101.144301, krapivsky2019free, wasak2019quantum,PhysRevB.101.075139,PhysRevLett.125.080402, PhysRevB.104.155431, PhysRevA.103.L060201, 10.21468/SciPostPhys.12.1.011, PhysRevB.103.L020302, Francesco_2021, Moca_2021}.
      
The onset of the Zeno effect is marked by slowing down of the dynamics of a quantum system, beyond a certain dissipation strength. In this work, we take one step further and we inspect whether  entanglement  quantifiers can expose the Zeno crossover, traditionally characterised by a smooth transition from facilitation (the heating regime) to suppression (the Zeno regime) of transport upon tuning the strength of local noise above the characteristic energy scale of the system as reported in \cite{PhysRevB.102.100301} and \cite{PhysRevB.101.144301}. Specifically, we consider an exactly solvable instance of the problem extending previous results on locally noise-driven free fermions~\cite{PhysRevB.102.100301}, to evaluate the noise averaged entanglement entropy of a bipartition of spins. The Gaussian nature of the problem allows us to average over several noise realisations, and to follow dynamics for long times and large system sizes. We show that the growth rate of the noise averaged entanglement entropy carries a hallmark of the Zeno crossover. 
In the conclusion we compare with recent analogous results on entanglement dynamics in free fermion chains subject to localised losses~\cite{10.21468/SciPostPhys.12.1.011}. 

\section{Theoretical framework}
\subsection{Model}
We study the quantum XY spin chain~\cite{fabio_quantum}
\begin{equation}
H_0  = \sum_{i = -L}^{L}-J(1+\gamma)\sigma_i^x\sigma_{i+1}^x - J(1-\gamma)\sigma_i^y\sigma_{i+1}^y - h\sigma_i^z,
\end{equation}
where $\sigma^{\alpha}_{i}, (\alpha = x,y,z)$  are the Pauli matrices for the $i^{th}$ spin along the direction $\alpha$, and $0 \leq \gamma \leq 1$. We set $J = 1$ for the rest of this work and measure all the energy scales with respect to it. We assume periodic boundary condition ($\sigma^{\alpha}_{i} = \sigma^{\alpha}_{2L+1+i}$) and work in the units of $\hbar = 1$.
The system is prepared in the ground state of $H_0$  and at $t = 0$ a noisy signal is suddenly turned on which drives the spin in the center, (dissipative impurity) i.e at the site $i=0$. The resulting dynamics is given by the time dependent Hamiltonian
\begin{equation} \label{SSE}
 H(t) = H_0 + \sqrt{\kappa}\eta(t)\sigma_{0}^{z}.
 \end{equation}
The noisy signal $\eta(t)$ is an Ornstein-Uhlenbeck (OU) process~\cite{gardiner2004quantum} with $\langle \eta(t) \rangle = 0$ and $\langle\eta(t_1)\eta(t_2)\rangle =  \exp{\left(-\frac{|t_1 - t_2|}{\tau}\right)}/\tau$. When $\tau \rightarrow 0$, $\eta(t)$ is equivalent to Gaussian white noise. In the following we will work in this limit, unless otherwise stated. The main exception is in \sref{sec:tau}, where we study the effects of finite $\tau$. The stochastic Schr\"odinger equation (SSE) with the Hamiltonian in \eref{SSE} then reads
\begin{equation}\label{SSE_1}
\partial_t |\psi^{\eta}(t) \rangle = -\rmi H(t) |\psi^{\eta}(t)\rangle,
\end{equation}
where $|\psi^{\eta}(t) \rangle$ represents a stochastic quantum trajectory for a given noise realisation. 
In the Gaussian white noise limit, the stochastic Schr\"odinger equation implies the following Lindbladian evolution~\cite{doi:10.1080/00107510601101934}
\begin{equation}\label{rhooo}
\partial_t\rho(t) = -\rmi[H_0,\rho(t)] + \frac{\kappa}{2}[\sigma^{z}_0,[\rho(t), \sigma^{z}_0]],
\end{equation}
where  $\rho(t) = \langle \rho^{\eta}(t) \rangle_{\eta} = \langle|\psi^{\eta}(t)\rangle \langle \psi^{\eta}(t)| \rangle_{\eta}$ is the noise averaged density matrix, and the symbol $\langle \cdots \rangle_{\eta}$ represents averaging over different noise realisations. In this paper, we study both the average properties of stochastic quantum trajectories captured by $\rho(t)$, and the properties of individual stochastic quantum trajectories encoded in $ \rho^{\eta}(t)$. We accomplish this by studying two types of observables: observables linear in $\rho^{\eta}(t)$ and observables non-linear in $\rho^{\eta} (t)$. Observables linear in $\rho^{\eta}(t)$, such as local and global magnetisation of spins, erase the information of an individual stochastic quantum trajectory upon averaging and end up capturing the properties of $\rho(t)$. However, non-linearity of the observables such as entanglement entropy of a block of spins in the interval $[l,m]$, generally allow them to retain the information of an individual stochastic quantum trajectory in spite of averaging~\cite{PhysRevB.105.064305, PhysRevLett.126.170602, 10.21468/SciPostPhys.7.2.024}. We explore the linear case in \sref{quantum XY chain with a noisy impurity}  and the non-linear case in \sref{sec:EE}.

To study the dynamics of this system we employ Jordan-Wigner transformation~\cite{fabio_quantum, Glen_2020} to map $H(t)$ to a system of spinless fermions:
\begin{equation}\label{JW}
\eqalign{
H(t) &= \sum_{i = -L}^{L-1} -2(c^{\dag}_{i}c_{i+1} + c^{\dag}_{i+1}c_{i})  - 2\gamma(c^{\dag}_{i}c^{\dag}_{i+1} + c_{i+1}c_{i})  \\& + 2U_p(c^{\dag}_{L}c_{-L} + c^{\dag}_{-L}c_{L} + \gamma c^{\dag}_{L}c^{\dag}_{-L} +  \gamma c_{-L}c_{L}) \\ & -h\sum_{i = -L}^{L}(c_{i}c^{\dag}_{i}-c^{\dag}_{i}c_{i}) + \sqrt{\kappa}\eta(t)(c_{0}c^{\dag}_{0}-c^{\dag}_{0}c_{0}).} 
\end{equation}
After the transformation, the periodic boundary condition of the spin chain map to a boundary term $2U_p(c^{\dag}_{L}c_{-L} + c^{\dag}_{-L}c_{L} + \gamma c^{\dag}_{L}c^{\dag}_{-L} +  \gamma c_{-L}c_{L})$, where $U_p = \Pi^{L}_{m=-L}\sigma^z_m$ is the parity operator. $U_{p}$ is known to be a symmetry of $H(t)$~\cite{fabio_quantum}, and this gives $H(t)$ a block diagonal structure in fermionic basis $(c_i,c^{\dag}_i)$. 
Each block is specified by the eigenvalue $\pm 1$ of $U_p$. The block corresponding to $+1$ is known as the even sector of $H(t)$ and similarly the block corresponding to  $-1$  is known as the odd sector of $H(t)$. Without loss of generality, we work in the even sector and hence the final Hamiltonian used for simulating the dynamics generated by \eref{SSE_1} is,
\begin{equation}
\eqalign{
H(t) &= \sum_{i = -L}^{L} -2(c^{\dag}_{i}c_{i+1} + c^{\dag}_{i+1}c_{i})  - 2\gamma(c^{\dag}_{i}c^{\dag}_{i+1} + c_{i+1}c_{i}) - h(c_{i}c^{\dag}_{i}-c^{\dag}_{i}c_{i})  \\& + \sqrt{\kappa}\eta(t)(c_{0}c^{\dag}_{0}-c^{\dag}_{0}c_{0}),}
\end{equation}
where $c_{2L+1+i} = -c_{i}$. As $H(t)$ has now reduced to a system of non-interacting spinless fermions, we can efficiently simulate its dynamics for large system sizes and for long times. In addition to numerical efficiency, the Jordan-Wigner transformation reveals the nature of the quasi-particle excitation of the Hamiltonian $H_0$. To elucidate this further, we use a vectorial notation
\begin{equation}
\bi{c} = \left(\begin{array}{cc}c_{-L}, \cdots, c_{L}, c^{\dag}_{-L}, \cdots, c^{\dag}_{L}\end{array}\right)^{T},
\end{equation} 
to express $H(t)$ in a more compact form
\begin{equation}
\label{eq_H_compact_matrix_form}
H(t) =  \bi{c}^{\dag}(M_0 + \sqrt{\kappa}\eta(t)\mathbb{L})\bi{c}.
\end{equation}
$M_0$ is the single particle Hamiltonian associated with the coherent part $H_0$ in $H(t)$ and $\mathbb{L}$ represents the impurity term $\sigma_0^{z}$ in $H(t)$,

\begin{equation}\label{single_particle_hamiltonian}
 M_0 = \left[\begin{array}{c|c}\alpha & -\beta \\ \hline \beta & -\alpha\end{array}\right], \quad \mathbb{L}  =  \left[\begin{array}{c|c}-\textbf{L} & 0  \\ \hline 0& \textbf{L}\end{array}\right],
 \end{equation}
whose matrix elements read 
\begin{eqnarray}
\alpha_{i,j} &=& -(\delta_{i,j+1} + \delta_{i,j-1} - \delta_{i,-L}\delta_{j,L} -  \delta_{i,L}\delta_{j,-L}) +h\delta_{i,j}, \\
\beta_{i,j} &=&  \gamma(\delta_{i,j-1} - \delta_{i,j+1} + \delta_{i,-L}\delta_{j,L} -  \delta_{i,L}\delta_{j,-L}), \\
\textbf{L}_{i,j} &=& \delta_{i,0}\delta_{j,0},
\end{eqnarray}
with $i,j \in [-L,L]$. The $\alpha_{ij}$ capture the hopping of fermions throughout the chain, while $\beta_{ij}$ are called anomalous densities, which lead to creation or destruction of two fermions out of the vacuum; finite $\beta_{ij}$ are at the root of lack of particle number conservation in fermionic systems~\cite{fabio_quantum, Glen_2020}. 

To identify the natural quasi-particle excitations of $H_0$, we use the standard procedure outlined as follows. First, we move to Fourier space using the following definition
\begin{equation} 
\tilde{c}_{q} = \frac{1}{\sqrt{(2L+1)}}\sum_{j = -L}^{L}c_je^{\rmi q j}, \quad q = \frac{2\pi}{2L+1}\left(n + \frac{1}{2}\right),
\end{equation} 
where $n = -L, -L+1, \cdots , L$, followed by a Bogoliubov transformation~\cite{fabio_quantum,Glen_2020,Calabrese_2012}
\begin{equation}\label{bogo_rotation}
\eqalign{
b_{q} &= u_q\tilde{c}_q - \rmi v_q\tilde{c}_{-q}^{\dag} \\ 
b_{-q}^{\dag} &= u_q\tilde{c}_{-q}^{\dag} - \rmi v_q\tilde{c}_{q}.}
\end{equation}
Here $u_q=\cos\left( {\theta_q}/{2}\right)$ and $v_q=\sin\left({\theta_q}/{2}\right)$, and $\theta_q$ is the Bogoliubov angle defined by
\begin{equation}\label{condition}
\tan\theta_q = \frac{2\gamma\sin q}{h -2 \cos q}. 
\end{equation}
Finally this overall procedure diagonalises $H_0$, such that it takes the following form in terms of Bogoliubov fermions ($b_{q}, b^{\dag}_{q}$)
\begin{equation}\label{natural_excitation}
H_0 = \sum_q \epsilon(q)\left(b^{\dag}_{q}b_{q} - \frac{1}{2}\right),
\end{equation}
where $\epsilon(q) = 4(\left(\cos(q)-\frac{h}{2}\right)^2 + \left(\gamma\sin(q)\right)^2)^{\frac{1}{2}}$. \Eref{natural_excitation} reveals Bogoliubov fermions as the natural quasi-particle excitations of $H_{0}$. 

\subsection{Structure of the dissipative impurity}

In this subsection, we look at the structure of the dissipative impurity to understand its effect on the dynamics. Under the Jordan-Wigner transformation followed by a Fourier transformation, the impurity term $\sigma_{0}^{z}$ takes the following form,
\begin{equation} \label{density}
\sigma^{z}_0 = 1 - \frac{2}{2L+1}\sum_{p,q}\tilde{c}_{p}^{\dag}\tilde{c}_{q}.
\end{equation}
In terms of Fourier transformed Jordan-Wigner fermions ($\tilde{c}_q, \tilde{c}^{\dag}_q$), the impurity term only contains local dephasing term ($\tilde{c}_{p}^{\dag}\tilde{c}_{q}$)~\cite{PhysRevB.102.100301}. Starting from the ground state of $H(t = 0^{-})$ for $\gamma = 0$, which is a uniform Fermi-sea, this local dephasing term redistributes the Fourier transformed Jordan-Wigner fermions already present in the Fermi-sea to the states with momentum larger than Fermi-momentum ($k_{F} = \cos^{-1}\left(\frac{h}{2}\right)$)~\cite{PhysRevB.102.100301}. For $\gamma = 0$, Fourier transformed Jordan-Wigner fermions are the same as the Bogoliubov fermions (cf. with equations \eref{bogo_rotation} and \eref{condition}) and hence serve as the natural quasi-particle excitations of $H_0$. However, when $\gamma > 0$, the action of the impurity on the Bogoliubov fermions changes dramatically as $\sigma_{0}^{z}$ takes the following form in terms Bogoliubov fermions,
\begin{equation}\label{magne}
\sigma^{z}_0 = 1 - \frac{2}{2L+1}\sum_{p,q}(-\rmi u_qv_pb_{-p}b_q +\rmi u_pv_qb_{p}^{\dag}b_{-q}^{\dag} + v_pv_qb_{-p}b_{-q}^{\dag} +u_qu_pb_{p}^{\dag}b_q).
\end{equation}
This representation of the impurity term $\sigma_{0}^{z}$ contains several other nontrivial terms like two body losses and pumps  ($b_{-p}b_q$ and $b_{p}^{\dag}b_{-q}^{\dag}$, respectively) in addition to local dephasing terms ($b_{p}^{\dag}b_q$ and $b_{-p}b_{-q}^{\dag}$). Since we start in the ground state of $H(t = 0^{-})$ in the even sector, which is a vacuum of Bogoliubov fermions~\cite{fabio_quantum, Glen_2020}, the two body pumps in the impurity term $\sigma^{z}_{0}$ generate dynamics by emitting pairs of quasi-particles, which are subjected to losses and dephasing during the evolution of the system.

\section{quantum XY chain with a noisy impurity}
\label{quantum XY chain with a noisy impurity}

\begin{figure*}[h!t]
\centering
 \includegraphics[scale = 0.4]{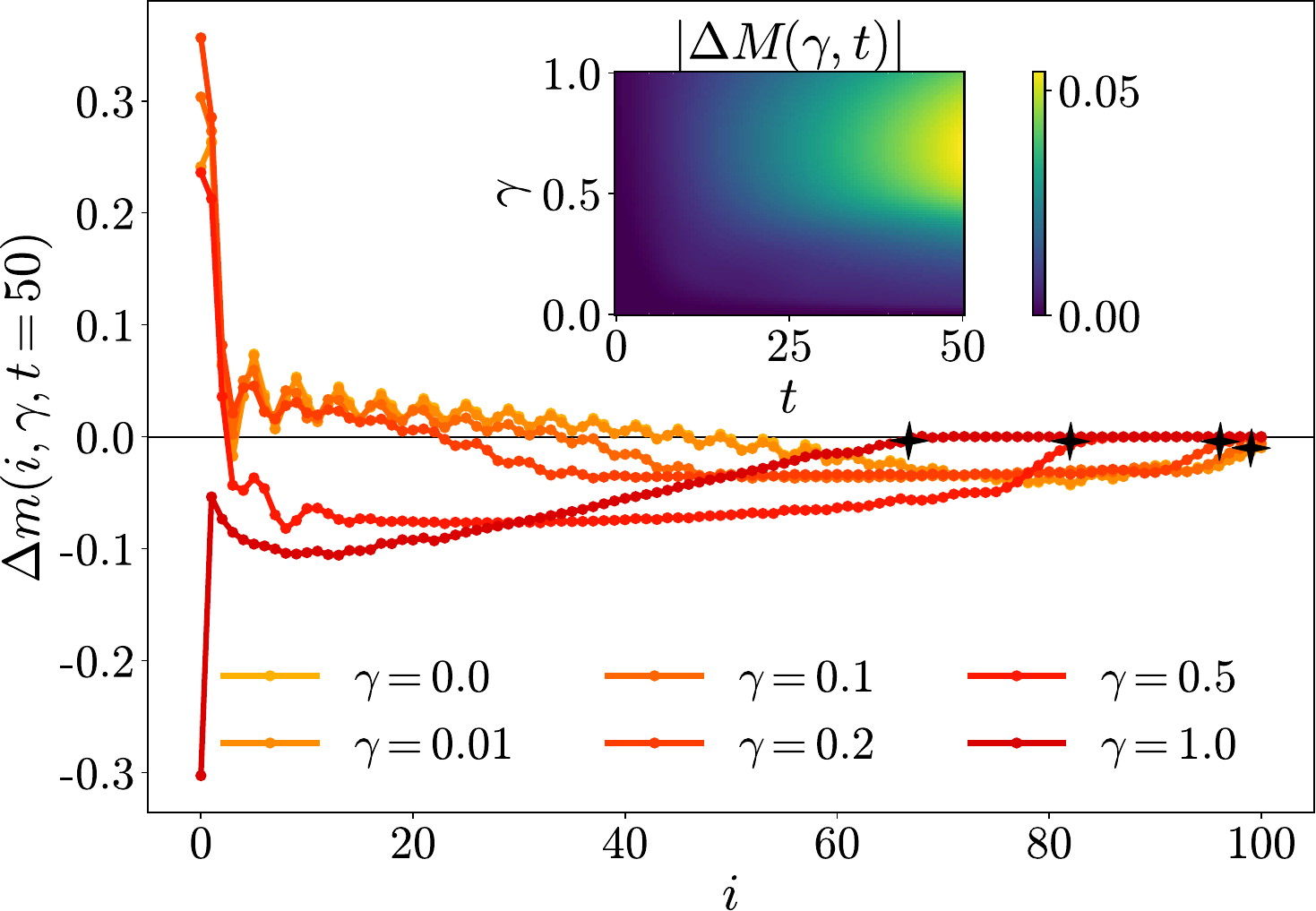}
\caption{ Local magnetisation with reference to its initial value $\Delta m(i,\gamma,t) = m(i,\gamma,t) - m(i,\gamma,0)$, at $t = 50$ and $i \in [0,L]$. We see a ballistically propagating magnetisation front (shown by black stars) emitted by the impurity, for all values of $\gamma$. The inset shows a density plot of absolute value of the global magnetisation ($M(\gamma,t)$) with reference to its initial value ($\Delta M(\gamma,t) = M(\gamma,t) - M(\gamma,0)$), for $\gamma \in [0,1]$ and $t \in [0,50]$. These results are obtained via numerical simulation of the adjoint master equation~\ref{single_particle_lindblad_equation}. We have chosen $L = 100$, $\kappa = 1$ and $h = \sqrt{2}$, for the results presented here.}
\label{fig1}
\end{figure*}

In this section we study the local magnetisation profile and the global magnetisation of the spin chain to further extend the results presented in reference~\cite{PhysRevB.102.100301}, before moving on to the central result of this paper in \sref{sec:EE}. First, we evaluate the dynamics of the local magnetisation $m(i,\gamma,t) = \Tr[\rho(t) \sigma_i^z]$, as a function of its position along the chain $i \in [-L,L]$ for different values of $\gamma$. We proceed by numerically simulating the adjoint master equation, which takes the following form (given that the jump operator is Hermitian)~\cite{Heinz_2010}
\begin{equation}\label{adjoint_equation}
\partial_t \langle A \rangle = \partial_t(\Tr(A\rho(t)) = \rmi\langle[H_0,A]\rangle + \frac{\kappa}{2}\langle[\sigma_0^{z},[A,\sigma_0^{z}]]\rangle.
\end{equation}
Using the equation \eref{adjoint_equation}, we evaluate the equations of motions for the two point correlators, $C_{i,j} =  \Tr[\rho(t)c^{\dag}_{i}c_{j}]$ and $F_{i,j} = \Tr[\rho(t)c^{\dag}_{i}c^{\dag}_{j}]$:
\begin{equation}\label{single_particle_lindblad_equation}
\fl\partial_t\slashed{C} = -\rmi[M_0,\slashed{C}] - 2\kappa\left[\begin{array}{c|c}-\textbf{L}C-C\textbf{L}+2\textbf{L}C\textbf{L} & \textbf{L}F^{\dag} + F^{\dag}\textbf{L} + 2\textbf{L}F^{\dag}\textbf{L} \\ \hline \textbf{L}F + F\textbf{L} + 2\textbf{L}F\textbf{L} & \textbf{L}C+C\textbf{L} - 2\textbf{L}C\textbf{L}\end{array}\right],
\end{equation}
where $\slashed{C}$ is a matrix which contains $C$ and $F$ matrices as follows
\begin{equation}\label{correlation_matrix}
\slashed{C} = \left[\begin{array}{c|c}\mathbb{I} - C  &  F^{\dag} \\ \hline F  & C\end{array}\right]. 
\end{equation}
In \eref{correlation_matrix}, $\mathbb{I}$ is the identity matrix of size $(2L+1) \times (2L+1)$. 
The local magnetisation is straightforwardly computed via $m(i,\gamma,t) = 1 - 2C_{i,i}(\gamma,t)$ and it is shown in \fref{fig1}.

For $\gamma > 0$, the dissipative impurity locally perturbs the chain and results in emission of pairs of quasi-particles which move with a velocity $\mathsf{v}(q) = \partial_q\epsilon(q)$, where $\epsilon(q) = 4(\left(\cos(q)-\frac{h}{2}\right)^2 + \left(\gamma\sin(q)\right)^2)^{\frac{1}{2}}$~\cite{Calabrese_2012,PhysRevB.84.165117}. The dynamics of these quasi-particles is limited by the light cone set by the maximum quasi-particle velocity $|\mathsf{v}^{max}| \equiv |\max\limits_{{q}} \mathsf{v}(q)|$ which is well reflected in \fref{fig1} as the ballistically propagating magnetisation front  occurring at $i = |\mathsf{v}^{max}|t$. This ballistically travelling magnetisation front separates region ($i > |\mathsf{v}^{max}|t$) where the local magnetisation has the same value as that of the initial state, from the region ($i < |\mathsf{v}^{max}|t$) where the local magnetisation value differs from that of the initial state (see \fref{fig1}). From \fref{fig1} we observe that at any given time the spatial extent of this light cone decreases as a function of $\gamma$. This follows from the quasi-particle velocity expression~\cite{Calabrese_2012, PhysRevB.84.165117},
\begin{equation}
\mathsf{v}(q) = -4\sin(q)\left(\frac{(1-\gamma^2)\cos(q) - \frac{h}{2}}{\sqrt{(\cos(q)-\frac{h}{2})^2 + \gamma^2\sin^2(q)}}\right).
\end{equation}
An increase in $\gamma$ leads to reduction of group velocity $|\mathsf{v}(q)|$ for any value of $q$. Therefore, the maximum quasi-particle velocity $|\mathsf{v}^{max}|$ decreases as $\gamma$ increases, confirming our earlier observation. Increasing $\gamma$ also changes the spatial oscillations that appear before the ballistically propagating magnetisation front. In \fref{fig1}, we generally observe reduction of oscillation amplitude and the spatial extent of the oscillations for $\gamma > 0$. 

When $\gamma = 0$, the system is symmetric under rotations along the $\hat{z}$-axis, thus the global magnetisation ($M(\gamma,t) = \sum_{i}m(i,\gamma,t)/(2L+1)$) along $\hat{z}$-axis remains conserved. However, this symmetry is broken for $\gamma \neq 0$ and hence to understand the effects of this symmetry breaking on the dynamics quantitatively, we calculate the difference between the global magnetisation and its initial value (see the inset in \fref{fig1}).  Despite the fact that, for $\gamma \neq 0$ the global magnetisation along $\hat{z}$ is not conserved, we observe that $M(\gamma,t)$ does not change appreciably as compared to its initial value $M(\gamma,0)$ up to time scales of the order of $1/\gamma$. The local magnetisation profile also reflects this behaviour. In \fref{fig1}, for small $\gamma$ (such as $\gamma = 0.01$), the local magnetisation profile  does not differ appreciably from the profile obtained  for $\gamma = 0$ upto the times reached during numerical simulation of the equation \eref{single_particle_lindblad_equation}.  However, as $\gamma$ increases, the local magnetisation profile deviates significantly from the profile obtained in the case of $\gamma = 0$. 


\section{Entanglement Entropy as a probe of the Zeno crossover}
\label{sec:EE}
\subsection{Observables and simulation methods}
 
We now turn our attention to the entanglement dynamics in presence of the dissipative impurity. We use the von Neumann entropy of the reduced density matrix to quantify entanglement entropy (EE)~\cite{RevModPhys.80.517} for the block of spins within the interval $[l,m]$
\begin{equation}
S^{\eta}_{ent}(t) = -\Tr[\rho^{\eta}_{[l,m]}\log_2\rho^{\eta}_{[l,m]}],
\end{equation}
where $\rho^{\eta}_{[l,m]}$ is the reduced density matrix $\Tr_{[l,m]^c}[\rho^{\eta}(t)]$ of the spins between site $l$ and site $m$, for a given noise realisation $\eta$. Here $\Tr_{[l,m]^c}$ refers to trace over the remainder of the chain. For the rest of the paper, we choose to study $S_{ent}(t) = \langle S^{\eta}_{ent}(t) \rangle_{\eta}$ over the EE of a block of spins between site $l$ and site $m$ in the noise averaged state 
\begin{equation}
\overline{S}_{ent}(t)=-\Tr[\rho_{[l,m]}\log_2\rho_{[l,m]}],
\end{equation}
where $\rho_{[l,m]} = \Tr_{[l,m]^c}[\rho(t)]$. Our choice stems from the fact that, the time evolution of $\rho(t)$ is governed by the equation \eref{rhooo} which does not preserve the purity of $\rho(t)$ at all times. Hence, $\overline{S}_{ent}(t)$ does not serve as a proper measure of entanglement~\cite{10.21468/SciPostPhys.7.2.024, 10.21468/SciPostPhys.12.1.011, PhysRevB.105.064305}.

\begin{figure}[h!t]
\centering
\includegraphics[scale = 0.4]{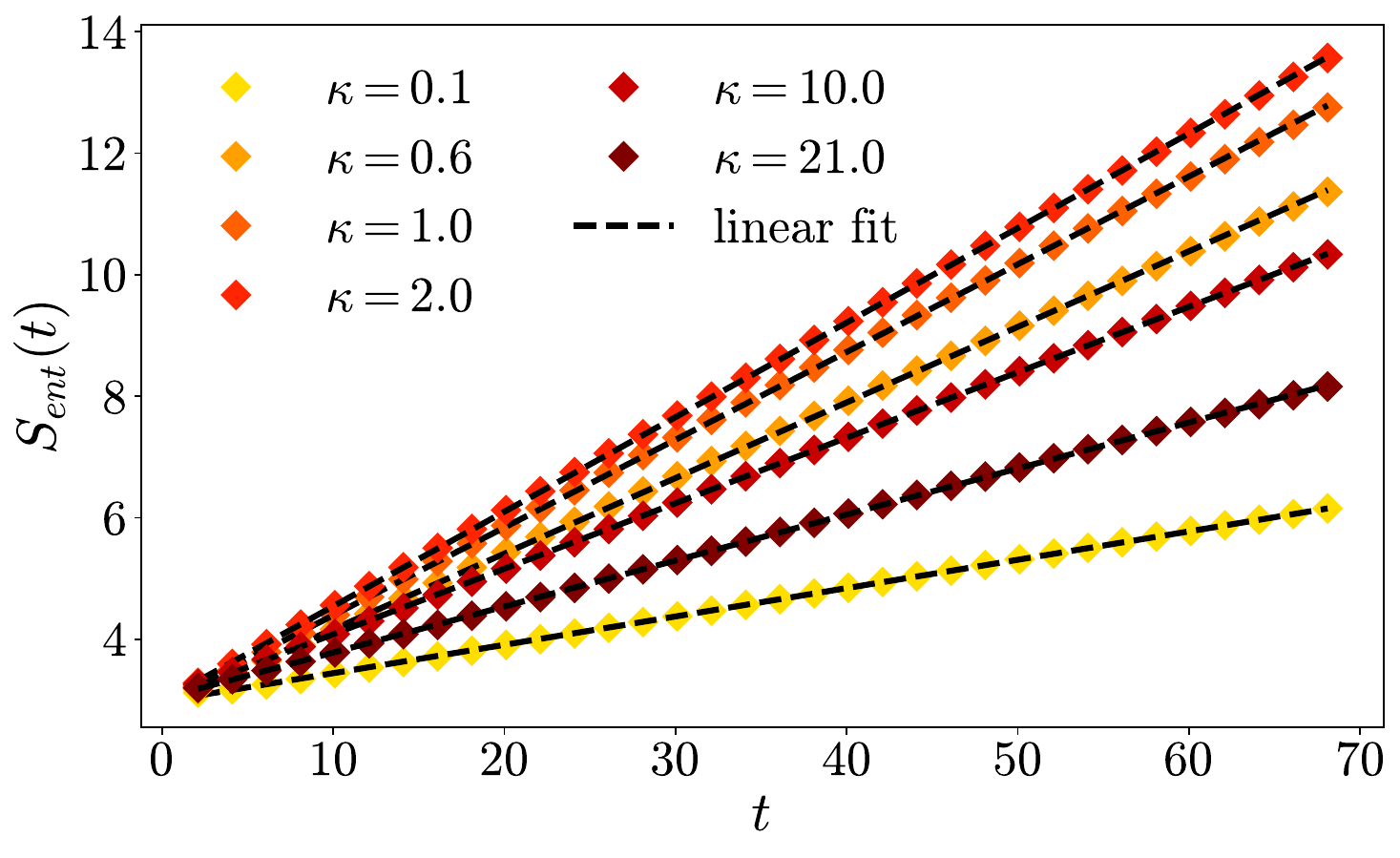}
\caption{The entanglement entropy  $S_{ent}(t) =  \langle S^{\eta}_{ent}(t) \rangle_{\eta}$ of the spins between $l = 0$ and $m = L$ for different values of $\kappa$. For each value of $\kappa$ we plot the linear fit (dashed black line). The results presented here are obtained via numerical simulation of the SSE~\eref{SSE_1} for $L = 160$, $\gamma = 0$, $\tau = 0$ and $h=\sqrt{2}$. They are averaged over $N_{\eta} = 1.1 \times 10^{4}$ noise realisations.} 
 \label{linear_ent}
\end{figure}

To compute  $S^{\eta}_{ent}(t)$, we use the fact that the stochastic quantum trajectory ($|\psi^{\eta}(t) \rangle$), is Gaussian and the SSE \eref{SSE_1} preserves the Gaussianity of the stochastic quantum trajectory~\cite{PhysRevB.102.100301,PhysRevB.105.064305}. Therefore the dynamics of each stochastic quantum trajectory can be obtained by computing the evolution of the two point correlation functions
\begin{equation}
\slashed{C}^{\eta}(t) = \left[\begin{array}{c|c}\mathbb{I}-C^{\eta}(t)& F^{\eta}(t)^{\dag} \\ \hline F^{\eta}(t) & C^{\eta}(t)\end{array}\right],
\end{equation}
where $C^{\eta}_{i,j}(t) = \Tr[\rho^{\eta}(t)c^{\dag}_{i}c_{j}]$ and $F^{\eta}_{i,j}(t) = \Tr[\rho^{\eta}(t)c^{\dag}_{i}c^{\dag}_{j}]$.

The numerical scheme that we use for obtaining $\slashed{C}^{\eta}(t)$ discretizes time in steps of $\delta t$. The $n^{th}$ instant is given as $t_n = (n-1)\delta t$ with $n = 1,2,...$ and $\delta t \ll $ min$\{h^{-1},\kappa^{-1}\}$. The correlation matrix $\slashed{C}^{\eta}_n = \slashed{C}^{\eta}(t_n)$ is computed from the correlation matrix $\slashed{C}^{\eta}_{n-1}$ by:
\begin{eqnarray}
&\slashed{C}^{\eta}_{n} = e^{-\rmi M^{\eta}\delta t}\slashed{C}^{\eta}_{n-1}e^{\rmi M^{\eta}\delta t},   \\
&M^{\eta} = M_0 + \sqrt{\kappa}\eta(t_{n-1})\mathbb{L}.
\end{eqnarray}
The operator $e^{-\rmi M^{\eta}\delta t}$ can be computed efficiently via Trotterization: 
\begin{equation*}
e^{-\rmi M^{\eta}\delta t} \approx e^{-\rmi \sqrt{\kappa}\eta(t_{n-1})\mathbb{L}\frac{\delta t}{2}}e^{-\rmi M_0 \delta t}e^{-\rmi \sqrt{\kappa}\eta(t_{n-1})\mathbb{L}\frac{\delta t}{2}},
\end{equation*}
where $M_0$ and $\mathbb{L}$ are defined by equation \eref{single_particle_hamiltonian}. In case of a Gaussian state, the entanglement entropy is given by~\cite{PhysRevB.94.245131,RevModPhys.80.517,lorenzo_quantum_2017}
\begin{equation}
S^{\eta}_{ent}(t) = -\sum_{p}[\lambda_p\log_2\lambda_p],
\end{equation}
where $\{\lambda_p\}$ are the eigenvalues of the matrix
\begin{equation}
\slashed{C}^{\eta}_{sub}(t) = \left[\begin{array}{c|c}(\mathbb{I}-C^{\eta}(t))_{i,j = l}^{m} & ((F^{\eta}(t))^{\dag})_{i,j = l}^{m} \\ \hline (F^{\eta}(t))_{i,j = l}^{m} & (C^{\eta}(t))_{i,j = l}^{m}\end{array}\right].
\end{equation}
The notation $(A)_{i,j = l}^{m}$ represents a matrix which contains all the elements of matrix $A$, between $l \leq i \leq m$ and $l \leq j \leq m$. 

\subsection{Entanglement dynamics}

\begin{figure}[h!t]
\centering
\includegraphics[scale = 0.4]{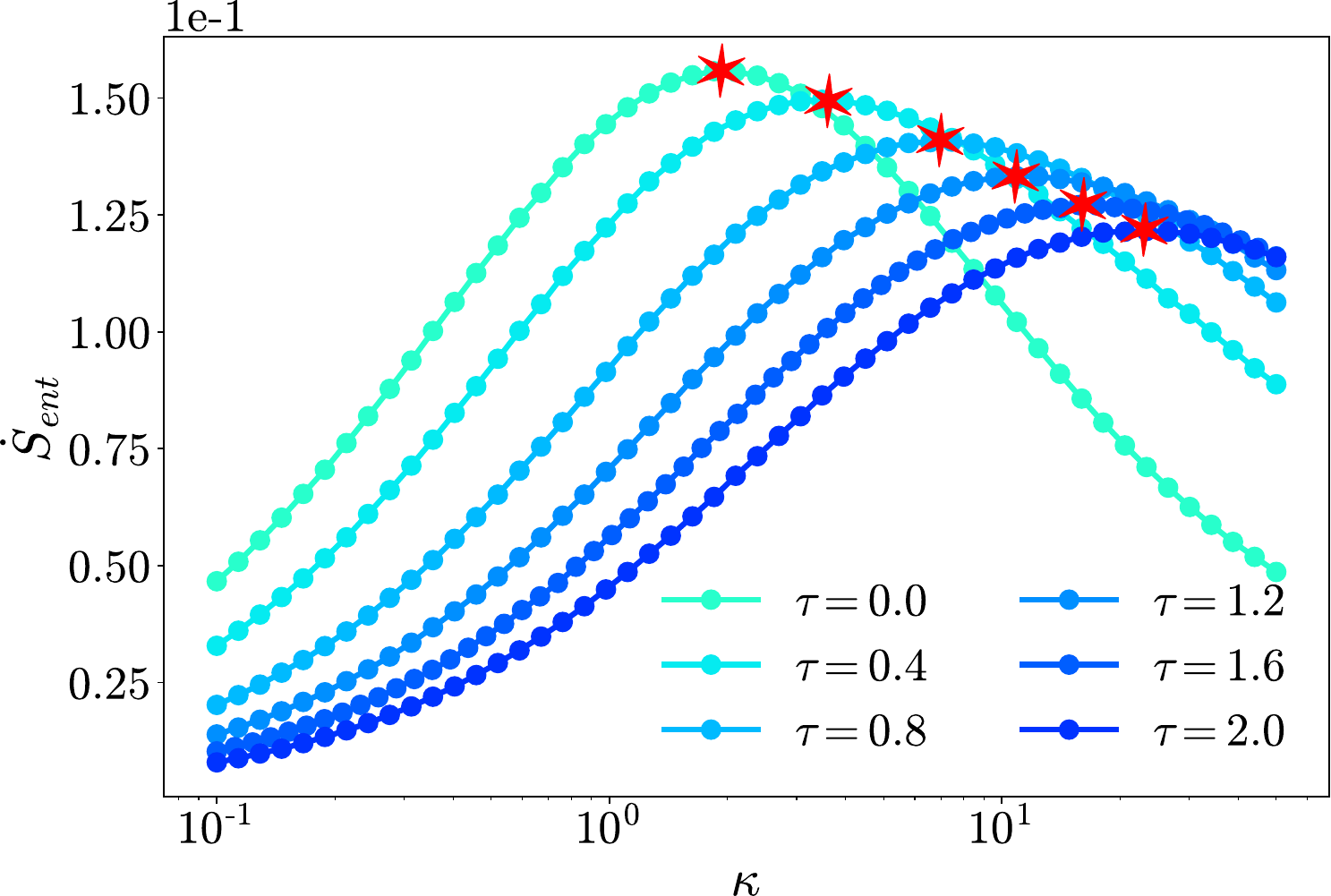}
\caption{Speed of entanglement spreading $\dot{S}_{ent} = \partial_t S_{ent}(t)$ of the spins between $l = 0$ and $m = L$ as a function of  $\kappa$. The red stars separate the heating regime from the Zeno regime for each value of $\tau$. The results presented here are obtained via numerical simulation of the SSE~\eref{SSE_1} for $L = 160$, $\gamma = 0$ and $h=\sqrt{2}$. They are averaged over $N_{\eta} = 1.1 \times 10^{4}$ noise realisations.}
\label{non_monotonous_ent}
\end{figure}

In this subsection we present two key features of EE dynamics for $\gamma = 0$. In particular we focus on the $S_{ent}(t) =  \langle S^{\eta}_{ent}(t) \rangle_{\eta}$ of the spins between site $l = 0$ and site $m = L$. We observe a linear growth of the $S_{ent}(t)$ (see \fref{linear_ent}) as expected, and  a non-monotonous behaviour of the $\dot{S}_{ent} = \partial_t S_{ent}(t)$ as a function of the coupling $\kappa$ to the dissipative impurity (see \fref{non_monotonous_ent}). Both observations can be qualitatively explained with the quasi-particle picture of entanglement spreading~\cite{calabrese2016quantum, 10.21468/SciPostPhys.12.1.011, PhysRevB.103.L020302,10.21468/SciPostPhys.7.2.024,Alba_2022,Alba_2022_1,PhysRevB.105.144305}.

The dissipative impurity scatters quasi-particles from the Fermi-sea and generates entangled pairs from reflected and transmitted quasi-particles~\cite{ 10.21468/SciPostPhys.12.1.011, PhysRevB.102.100301}. These entangled quasi-particles travel in opposite directions and entangle any two spatial sites at which they arrive simultaneously; this leads to the usual linear growth in time~\cite{calabrese2016quantum, 10.21468/SciPostPhys.12.1.011, 10.21468/SciPostPhys.7.2.024} of the $S_{ent}(t)$ (see \fref{linear_ent}). In the thermodynamic limit the light cone that originates at the driven site will take an infinite time to reach the edge of the system, hence the $S_{ent}(t)$ of the bipartition we are studying will increase unboundedly. 

The non-monotonous behaviour of the $\dot{S}_{ent}$ reflects the system's transition from the heating regime to the Zeno regime. In the heating regime, the system heats up as $\kappa$ increases while in the Zeno regime, the system's dynamics slow down with an increase in $\kappa$ as a consequence of the Zeno effect. Deep within the heating regime ($\kappa \ll 2$), we observe a very few scattering events as the coupling to the impurity is weak. This can be seen in \fref{fraction_of_scattered_quasi-particles}, where we show the fraction of quasi-particles scattered by the impurity,
\begin{equation}
f(t) = \frac{\sum^{k_F}_{q = -k_F} (\langle n^{\eta}_q(0) \rangle_{\eta} - \langle n_{q}^{\eta}(t) \rangle_{\eta})}{\sum^{k_F}_{q = -k_F} \langle n^{\eta}_q(0) \rangle_{\eta}}, \quad n_q^{\eta}(t) = \Tr[\rho^{\eta}(t)\tilde{c}_q^{\dag}\tilde{c}_q].
\end{equation}
Fewer scattering events lead to generation of a very few entangled quasi-particle pairs, which can take part in entangling the left and the right halves of the system. As $\kappa$ increases, the number of scattering events goes up, leading to an increase in the number of entangled quasi-particles pairs. Hence we observe an increase in $\dot{S}_{ent}$. However, at $\kappa \sim 2$, we observe a peak in $\dot{S}_{ent}$ which signals the system's transition from the heating regime to the Zeno regime. In the Zeno regime, the number of scattering events decrease, inspite of the strong coupling to the impurity (see \fref{fraction_of_scattered_quasi-particles}) as a consequence of the Zeno effect. This leads to a decline in number of entangled quasi-particle pairs as $\kappa$ increases. We also observe that, unlike in the heating regime where the entangled quasi-particle pairs easily spread across the system, in the Zeno regime, most of them remain trapped near the impurity site ($i = 0$). This is reflected in \fref{effect_of_tau_on_local_mag} where we plot the local magnetisation profile for different values of $\kappa$. For $\kappa = 21.0$, the local magnetisation profile strongly peaks near the impurity site as compared to $\kappa = 0.2$ in which case the local magnetisation profile spreads out evenly across all the sites.
 
Both, the decrease in number of entangled pairs of quasi-particles and the increase in the effectiveness of trapping of entangled pairs near the impurity site in the Zeno regime, lead to decline of $\dot{S}_{ent}$ as $\kappa$ increases which was recently dubbed as ``Zeno entanglement death" in the reference~\cite{10.21468/SciPostPhys.12.1.011}. Based on the discussion in this subsection, we expect the noise averaged entanglement entropy to take the following form which was also reported in reference~\cite{10.21468/SciPostPhys.12.1.011}
\begin{equation}\label{ansatz}
S_{ent}(t) = G(\kappa)\int^{k_F}_{-k_F} \frac{dk}{2\pi} S(k) \mathsf{v}_{k}t + S_0,
\end{equation}
where $S(k)$ is the entanglement entropy contribution from a single pair of entangled quasi-particles moving with a a wave vector $k$, $S_0$ is the initial entanglement entropy of the block of spins in the interval $[0,L]$ and $G(\kappa)$ is the function which reproduces the Zeno crossover. After comparing the equation \eref{ansatz} with the results in \fref{non_monotonous_ent}, we find that $G(\kappa)$ transitions from $A\kappa^{\alpha}$ to $B\kappa^{-\beta}$ as the system crosses over from the heating regime to the Zeno regime, where $\alpha$ and $\beta$ are both positive real exponents.

\subsection{Effect of a finite noise correlation time}
\label{sec:tau}

In this subsection we discuss the effects of finite correlation time $\tau$ of the noise on the $\dot{S}_{ent}$ in the heating regime and the Zeno regime. Deep inside the heating regime ($\kappa \ll 2$), the $\dot{S}_{ent}$ reduces as $\tau$ increases  (see \fref{non_monotonous_ent}). This happens as the fraction of quasi-particles scattered by the impurity goes down with increasing $\tau$ and it results in generation of lesser number of entangled quasi-particle pairs as compared to $\tau= 0$ case. 

\begin{figure}[h!t]
\begin{center}
\includegraphics[scale = 0.4]{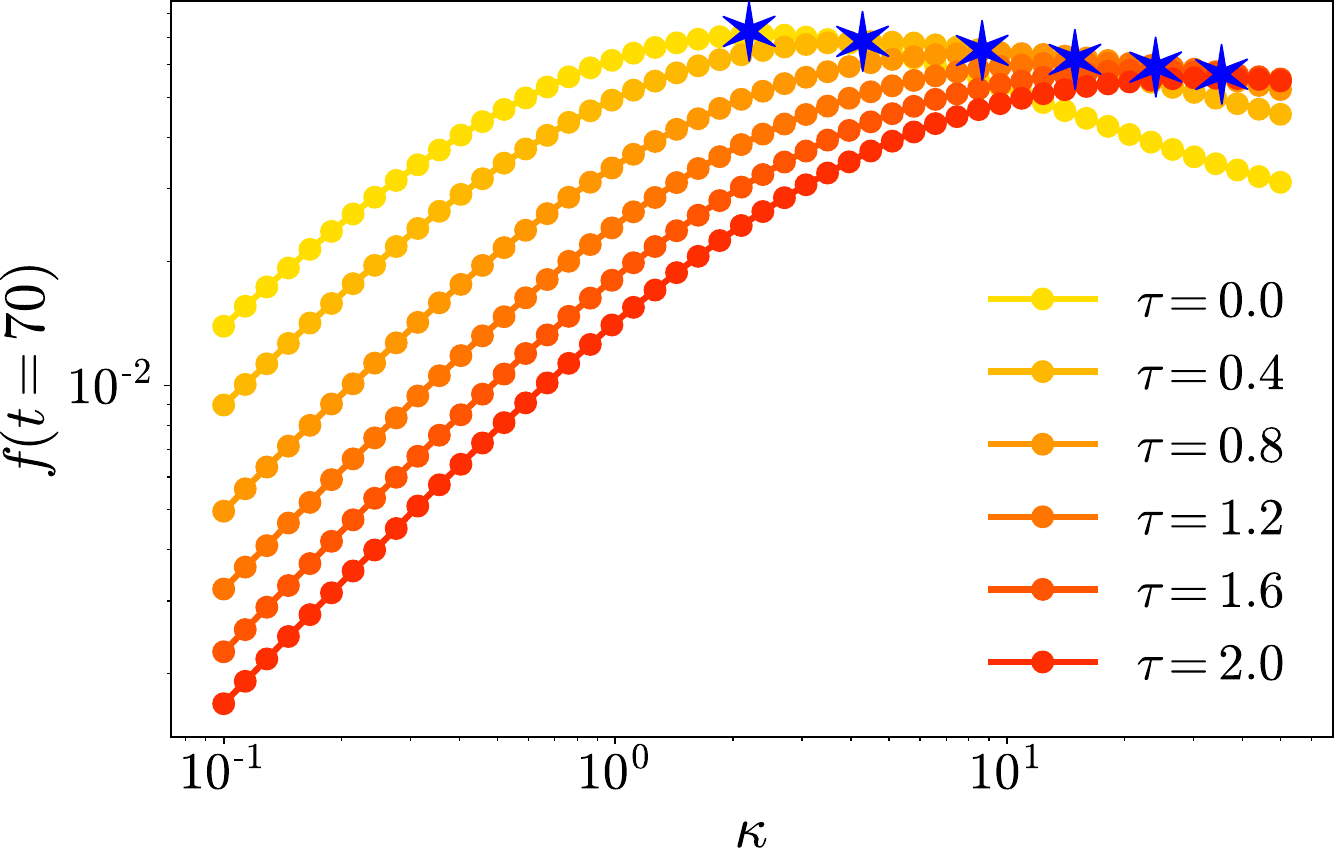}
 \caption{Fraction of quasi-particles scattered $f(t)$ at $t = 70$  as a function of $\kappa$. The blue stars separate the heating regime from the Zeno regime for each value of $\tau$. Deep in the heating regime ($\kappa \ll 2$), $f(t=70)$ grows algebraically with $\kappa$, while in the Zeno regime it decreases. These results were computed using the same parameters as discussed in \fref{non_monotonous_ent}.}
\label{fraction_of_scattered_quasi-particles}
\end{center}
\end{figure}
 
\begin{figure}[h!t]
\centering
 \includegraphics[scale = 0.4]{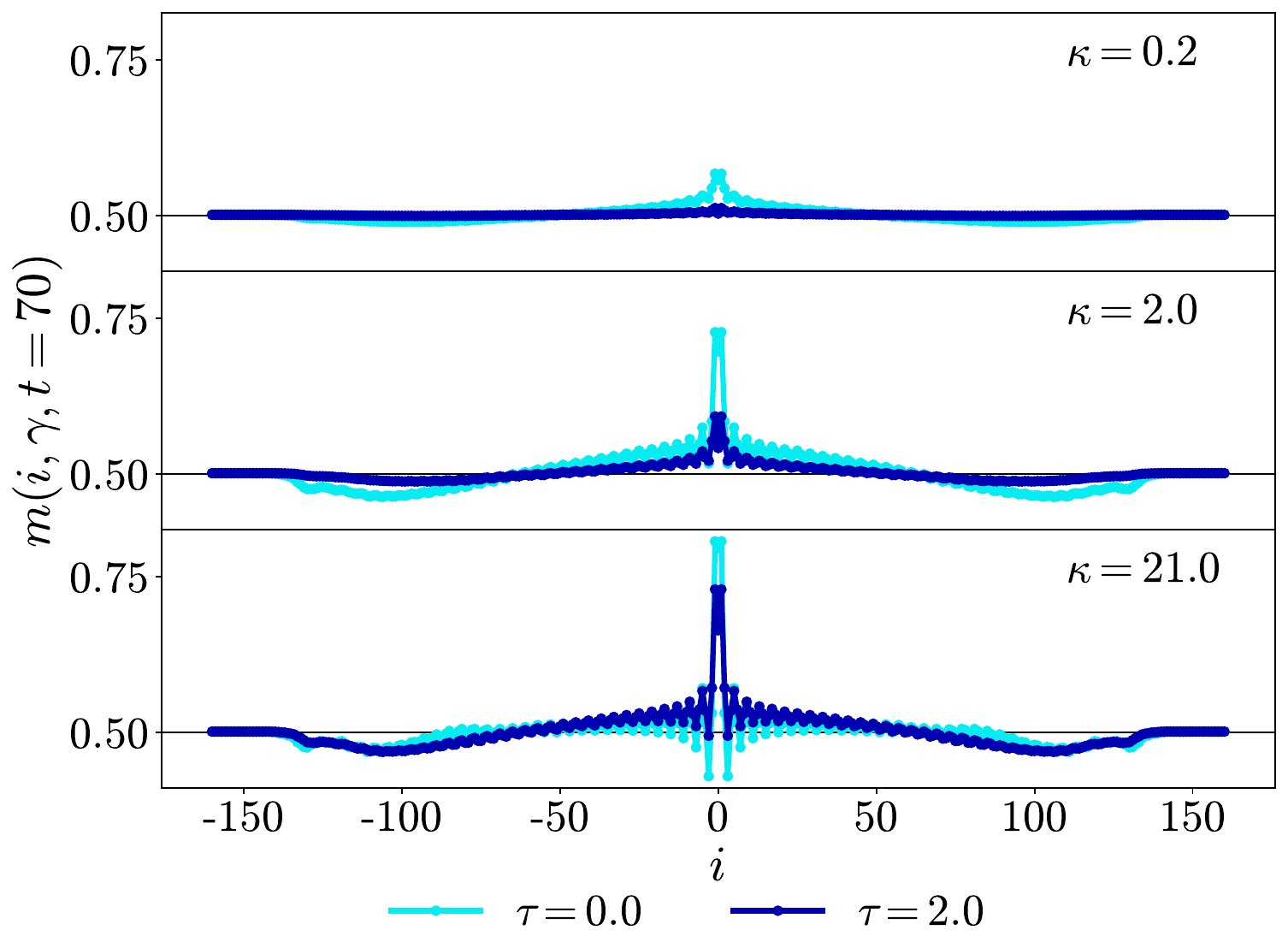}
 \caption{Local magnetisation profile $m(i, \gamma, t) = \langle m^{\eta}(i,\gamma,t) \rangle_{\eta}$ at $t = 70$ and $i \in [-L,L]$, for $\tau = 0$ and $\tau = 2$. 
 These results were computed using the same parameters as discussed in \fref{non_monotonous_ent}.}
 \label{effect_of_tau_on_local_mag}
\end{figure}

Deep in the Zeno regime ($\kappa \gg 2$) we observe the exact opposite: the suppression of $\dot{S}_{ent}$ reduces with increasing $\tau$. Indeed, as $\tau$ increases the trapping of entangled quasi-particles pairs near the impurity site $(i = 0)$ becomes less effective (cf. with \fref{effect_of_tau_on_local_mag}) and also the fraction of quasi-particles scattered by the impurity increases as compared to the $\tau = 0$ case (see \fref{fraction_of_scattered_quasi-particles}). This  allows more entangled quasi-particle pairs to participate in entangling the sub-system [$0,L$] with its complement, thus we observe larger values of $\dot{S}_{ent}$ as $\tau$ increases. In other words, the dissipative impurity dynamics slow down at larger $\tau$ and  the Zeno effect is expected to attenuate; this leads to a shift of the Zeno crossover to larger values of $\kappa$ (cf. with \fref{non_monotonous_ent}). 

We also observe that increasing $\tau$ leads to slower decline in $\dot{S}_{ent}$ as a function of $\kappa$ in the Zeno regime (see \fref{non_monotonous_ent}). This suggested a possibility of existence of a $\tau$ value at which there will be no decline in $\dot{S}_{ent}$ as a function of $\kappa$, which would have served as an indicator for the disappearance of the Zeno effect. However we carried out numerical simulations for larger values of $\tau$ and observed that the Zeno effect never disappears as a result of increasing $\tau$; the only net effect of a dissipative impurity with finite correlation time is to extend the heating regime to larger $\kappa$.



\section{Conclusions}

Recently, the entanglement dynamics in a free-fermion chain subjected to localised losses was studied thoroughly in reference~\cite{10.21468/SciPostPhys.12.1.011}. In that case, entanglement entropies do not serve as a proper entanglement measure since the system does not remain in a pure state, because of the non-unitary dynamics ensued by localised losses. Here we investigated the dynamics of a quantum XY spin chain driven by a local Ornstein-Uhlenbeck (OU) process with a finite noise correlation time ($\tau$). In spite of the driving by an OU process, the system always stays in a pure state for all noise realisations and hence entanglement entropy serves as a proper entanglement measure. In particular, our simulations reveal that, the noise averaged entanglement entropy of a bipartition of spins grows linearly as a function of time similar to the linear growth of logarithmic negativity observed in the case of the free-fermions with localised losses. Moreover, we see that the speed of spreading of the noise averaged entanglement entropy captures the transition from the heating regime to the Zeno regime, thus probing the Zeno crossover. We also observe that, increasing $\tau$ prolongs the heating regime and attenuates the Zeno suppression of noise averaged entanglement spreading in the Zeno regime. All our results are qualitatively explained by employing the quasi-particle picture of entanglement spreading similar to the entanglement spreading mechanism reported in~\cite{10.21468/SciPostPhys.12.1.011, PhysRevB.103.L020302, Alba_2022,Alba_2022_1,PhysRevB.105.144305}. 

It could be interesting to apply methods for the exact solution of boundary driven problems~\cite{vzunkovivc2010explicit, vznidarivc2011solvable, prosen2015matrix, vznidarivc2015relaxation, berdanier2019universal,Bu_a_2020,PhysRevB.105.054303} in order to extend our results to interacting versions of the free fermion model considered here. A relevant question is the possibility to morph the Zeno effect from a crossover into a sharp transition in the presence of interactions, and inspect the impact of a driving  noise with finite correlation time on such transition~\cite{PhysRevB.101.144301, PhysRevLett.122.040402}. Our current efforts are directed in this direction. 

\ack
We thank B. Buca for useful discussions. JM gratefully acknowledges P. Dolgirev, D. Sels and E. Demler for  collaborations on related topics. AC thanks R. Singh for helpful discussions and for the computational resources. The support and the resources provided by ‘PARAM Shivay Facility’ under the National Supercomputing Mission, Government of India at the Indian Institute of Technology, Varanasi are gratefully acknowledged.

\section*{References}
\bibliographystyle{iopart-num}
\bibliography{references}

\providecommand{\newblock}{}
\begin{thebibliography}{10}
\expandafter\ifx\csname url\endcsname\relax
  \def\url#1{{\tt #1}}\fi
\expandafter\ifx\csname urlprefix\endcsname\relax\def\urlprefix{URL }\fi
\providecommand{\eprint}[2][]{\url{#2}}

\bibitem{affleck_quantum_2009}
Affleck I 2009 Quantum {Impurity} {Problems} in {Condensed} {Matter} {Physics}
  arXiv: \href{http://arxiv.org/abs/0809.3474}{0809.3474}[cond-mat]

\bibitem{mahan2013many}
Mahan G~D 2013 {\em Many-particle physics\/} (Springer Science \& Business
  Media)

\bibitem{Ott3}
Zezyulin D~A, Konotop V~V, Barontini G and Ott H 2012 {\em Phys. Rev. Lett.\/}
  {\bf
  \href{https://link.aps.org/doi/10.1103/PhysRevLett.109.020405}{109}}(\href{https://link.aps.org/doi/10.1103/PhysRevLett.109.020405}{2})
  \href{https://link.aps.org/doi/10.1103/PhysRevLett.109.020405}{020405}

\bibitem{Ott4}
Barontini G, Labouvie R, Stubenrauch F, Vogler A, Guarrera V and Ott H 2013
  {\em Phys. Rev. Lett.\/} {\bf
  \href{https://link.aps.org/doi/10.1103/PhysRevLett.110.035302}{110}}(\href{https://link.aps.org/doi/10.1103/PhysRevLett.110.035302}{3})
  \href{https://link.aps.org/doi/10.1103/PhysRevLett.110.035302}{035302}

\bibitem{misra1977zeno}
Misra B and Sudarshan E~G 1977 {\em J. Math. Phys.\/} {\bf
  \href{https://doi.org/10.1063/1.523304}{18}}
  \href{https://doi.org/10.1063/1.523304}{756--763}

\bibitem{itano1990quantum}
Itano W~M, Heinzen D~J, Bollinger J and Wineland D 1990 {\em Phys. Rev. A\/}
  {\bf \href{https://doi.org/10.1103/PhysRevA.41.2295}{41}}
  \href{https://doi.org/10.1103/PhysRevA.41.2295}{2295}

\bibitem{facchi2002quantum}
Facchi P and Pascazio S 2002 {\em Phys. Rev. Lett.\/} {\bf
  \href{https://doi.org/10.1103/PhysRevLett.89.080401}{89}}
  \href{https://doi.org/10.1103/PhysRevLett.89.080401}{080401}

\bibitem{kofman1996quantum}
Kofman A and Kurizki G 1996 {\em Phys. Rev. A\/} {\bf
  \href{https://doi.org/10.1103/PhysRevA.54.R3750}{54}}
  \href{https://doi.org/10.1103/PhysRevA.54.R3750}{R3750}

\bibitem{kofman2001zeno}
Kofman A, Kurizki G and Opatrn{\`y} T 2001 {\em Phys. Rev. A\/} {\bf
  \href{https://doi.org/10.1103/PhysRevA.63.042108}{63}}
  \href{https://doi.org/10.1103/PhysRevA.63.042108}{042108}

\bibitem{kofman2000acceleration}
Kofman A and Kurizki G 2000 {\em Nature\/} {\bf
  \href{https://doi.org/10.1038/35014537}{405}}
  \href{https://doi.org/10.1038/35014537}{546--550}

\bibitem{PhysRevB.98.205136}
Li Y, Chen X and Fisher M~P~A 2018 {\em Phys. Rev. B\/} {\bf
  \href{https://link.aps.org/doi/10.1103/PhysRevB.98.205136}{98}}(\href{https://link.aps.org/doi/10.1103/PhysRevB.98.205136}{20})
  \href{https://link.aps.org/doi/10.1103/PhysRevB.98.205136}{205136}

\bibitem{PhysRevLett.126.170602}
Alberton O, Buchhold M and Diehl S 2021 {\em Phys. Rev. Lett.\/} {\bf
  \href{https://link.aps.org/doi/10.1103/PhysRevLett.126.170602}{126}}(\href{https://link.aps.org/doi/10.1103/PhysRevLett.126.170602}{17})
  \href{https://link.aps.org/doi/10.1103/PhysRevLett.126.170602}{170602}

\bibitem{PhysRevLett.122.040402}
Fr\"oml H, Chiocchetta A, Kollath C and Diehl S 2019 {\em Phys. Rev. Lett.\/}
  {\bf
  \href{https://link.aps.org/doi/10.1103/PhysRevLett.122.040402}{122}}(\href{https://link.aps.org/doi/10.1103/PhysRevLett.122.040402}{4})
  \href{https://link.aps.org/doi/10.1103/PhysRevLett.122.040402}{040402}

\bibitem{lebrat2019quantized}
Lebrat M, H{\"a}usler S, Fabritius P, Husmann D, Corman L and Esslinger T 2019
  {\em Phys. Rev. Lett.\/} {\bf
  \href{https://doi.org/10.1103/PhysRevLett.123.193605}{123}}
  \href{https://doi.org/10.1103/PhysRevLett.123.193605}{193605}

\bibitem{corman2019quantized}
Corman L, Fabritius P, H{\"a}usler S, Mohan J, Dogra L~H, Husmann D, Lebrat M
  and Esslinger T 2019 {\em Phys. Rev. A\/} {\bf
  \href{https://doi.org/10.1103/PhysRevA.100.053605}{100}}
  \href{https://doi.org/10.1103/PhysRevA.100.053605}{053605}

\bibitem{tonielli2019orthogonality}
Tonielli F, Fazio R, Diehl S and Marino J 2019 {\em Phys. Rev. Lett.\/} {\bf
  \href{https://doi.org/10.1103/PhysRevLett.122.040604}{122}}
  \href{https://doi.org/10.1103/PhysRevLett.122.040604}{040604}

\bibitem{schiro2019quantum}
Schiro M and Scarlatella O 2019 {\em The Journal of chemical physics\/} {\bf
  \href{https://doi.org/10.1063/1.5100157}{151}}
  \href{https://doi.org/10.1063/1.5100157}{044102} ISSN 0021-9606

\bibitem{tonielli2020ramsey}
Tonielli F, Chakraborty N, Grusdt F and Marino J 2020 {\em Phys. Rev.
  Research\/} {\bf
  \href{https://link.aps.org/doi/10.1103/PhysRevResearch.2.032003}{2}}(\href{https://link.aps.org/doi/10.1103/PhysRevResearch.2.032003}{3})
  \href{https://link.aps.org/doi/10.1103/PhysRevResearch.2.032003}{032003}

\bibitem{PhysRevA.103.043304}
Baals C, Moreno A~G, Jiang J, Benary J and Ott H 2021 {\em Phys. Rev. A\/} {\bf
  \href{https://link.aps.org/doi/10.1103/PhysRevA.103.043304}{103}}(\href{https://link.aps.org/doi/10.1103/PhysRevA.103.043304}{4})
  \href{https://link.aps.org/doi/10.1103/PhysRevA.103.043304}{043304}

\bibitem{Biella2021manybodyquantumzeno}
Biella A and Schir{\'{o}} M 2021 {\em {Quantum}\/} {\bf
  \href{https://doi.org/10.22331/q-2021-08-19-528}{5}}
  \href{https://doi.org/10.22331/q--2021--08--19--528}{528} ISSN 2521-327X

\bibitem{sartori2015spin}
Sartori A, Marino J, Stringari S and Recati A 2015 {\em New Journal of
  Physics\/} {\bf
  \href{https://iopscience.iop.org/article/10.1088/1367-2630/17/9/093036/meta}{17}}
  \href{https://iopscience.iop.org/article/10.1088/1367--2630/17/9/093036/meta}{093036}

\bibitem{yoshimura2020non}
Yoshimura T, Bidzhiev K and Saleur H 2020 {\em Phys. Rev. B\/} {\bf
  \href{https://link.aps.org/doi/10.1103/PhysRevB.102.125124}{102}}(\href{https://link.aps.org/doi/10.1103/PhysRevB.102.125124}{12})
  \href{https://link.aps.org/doi/10.1103/PhysRevB.102.125124}{125124}

\bibitem{PhysRevResearch.3.L032013}
Khedri A, \ifmmode~\check{S}\else \v{S}\fi{}trkalj A, Chiocchetta A and
  Zilberberg O 2021 {\em Phys. Rev. Research\/} {\bf
  \href{https://link.aps.org/doi/10.1103/PhysRevResearch.3.L032013}{3}}(\href{https://link.aps.org/doi/10.1103/PhysRevResearch.3.L032013}{3})
  \href{https://link.aps.org/doi/10.1103/PhysRevResearch.3.L032013}{L032013}

\bibitem{PhysRevLett.126.120603}
Maimbourg T, Basko D~M, Holzmann M and Rosso A 2021 {\em Phys. Rev. Lett.\/}
  {\bf
  \href{https://link.aps.org/doi/10.1103/PhysRevLett.126.120603}{126}}(\href{https://link.aps.org/doi/10.1103/PhysRevLett.126.120603}{12})
  \href{https://link.aps.org/doi/10.1103/PhysRevLett.126.120603}{120603}

\bibitem{vzunkovivc2010explicit}
{\v{Z}}unkovi{\v{c}} B and Prosen T 2010 {\em J. Stat. Mech.: Theory Exp.\/}
  {\bf 2010}
  \href{https://iopscience.iop.org/article/10.1088/1742--5468/2010/08/P08016}{P08016}

\bibitem{vznidarivc2011solvable}
{\v{Z}}nidari{\v{c}} M 2011 {\em Phys. Rev. E\/} {\bf
  \href{https://doi.org/10.1103/PhysRevE.83.011108}{83}}
  \href{https://doi.org/10.1103/PhysRevE.83.011108}{011108}

\bibitem{prosen2015matrix}
Prosen T 2015 {\em J. Phys. A\/} {\bf
  \href{https://iopscience.iop.org/article/10.1088/1751-8113/48/37/373001}{48}}
  \href{https://iopscience.iop.org/article/10.1088/1751--8113/48/37/373001}{373001}

\bibitem{vznidarivc2015relaxation}
{\v{Z}}nidari{\v{c}} M 2015 {\em Phys. Rev. E\/} {\bf
  \href{https://doi.org/10.1103/PhysRevE.92.042143}{92}}
  \href{https://doi.org/10.1103/PhysRevE.92.042143}{042143}

\bibitem{berdanier2019universal}
Berdanier W, Marino J and Altman E 2019 {\em Phys. Rev. Lett.\/} {\bf
  \href{https://link.aps.org/doi/10.1103/PhysRevLett.123.230604}{123}}(\href{https://link.aps.org/doi/10.1103/PhysRevLett.123.230604}{23})
  \href{https://link.aps.org/doi/10.1103/PhysRevLett.123.230604}{230604}

\bibitem{Bu_a_2020}
Bu{\v{c}}a B, Booker C, Medenjak M and Jaksch D 2020 {\em New Journal of
  Physics\/} {\bf 22} \href{https://doi.org/10.1088/1367--2630/abd124}{123040}

\bibitem{PhysRevB.105.054303}
Alba V and Carollo F 2022 {\em Phys. Rev. B\/} {\bf
  \href{https://link.aps.org/doi/10.1103/PhysRevB.105.054303}{105}}(\href{https://link.aps.org/doi/10.1103/PhysRevB.105.054303}{5})
  \href{https://link.aps.org/doi/10.1103/PhysRevB.105.054303}{054303}

\bibitem{PhysRevB.101.144301}
Fr\"oml H, Muckel C, Kollath C, Chiocchetta A and Diehl S 2020 {\em Phys. Rev.
  B\/} {\bf
  \href{https://link.aps.org/doi/10.1103/PhysRevB.101.144301}{101}}(\href{https://link.aps.org/doi/10.1103/PhysRevB.101.144301}{14})
  \href{https://link.aps.org/doi/10.1103/PhysRevB.101.144301}{144301}

\bibitem{krapivsky2019free}
Krapivsky P, Mallick K and Sels D 2019 {\em J. Stat. Mech.: Theory Exp.\/} {\bf
  2019} \href{https://doi.org/10.1088/1742--5468/ab4e8e}{113108}

\bibitem{wasak2019quantum}
Wasak T, Schmidt R and Piazza F 2021 {\em Phys. Rev. Research\/} {\bf
  \href{https://link.aps.org/doi/10.1103/PhysRevResearch.3.013086}{3}}(\href{https://link.aps.org/doi/10.1103/PhysRevResearch.3.013086}{1})
  \href{https://link.aps.org/doi/10.1103/PhysRevResearch.3.013086}{013086}

\bibitem{PhysRevB.101.075139}
Wolff S, Sheikhan A, Diehl S and Kollath C 2020 {\em Phys. Rev. B\/} {\bf
  \href{https://link.aps.org/doi/10.1103/PhysRevB.101.075139}{101}}(\href{https://link.aps.org/doi/10.1103/PhysRevB.101.075139}{7})
  \href{https://link.aps.org/doi/10.1103/PhysRevB.101.075139}{075139}

\bibitem{PhysRevLett.125.080402}
Mitchison M~T, Fogarty T, Guarnieri G, Campbell S, Busch T and Goold J 2020
  {\em Phys. Rev. Lett.\/} {\bf
  \href{https://link.aps.org/doi/10.1103/PhysRevLett.125.080402}{125}}(\href{https://link.aps.org/doi/10.1103/PhysRevLett.125.080402}{8})
  \href{https://link.aps.org/doi/10.1103/PhysRevLett.125.080402}{080402}

\bibitem{PhysRevB.104.155431}
M\"uller T, Gievers M, Fr\"oml H, Diehl S and Chiocchetta A 2021 {\em Phys.
  Rev. B\/} {\bf
  \href{https://link.aps.org/doi/10.1103/PhysRevB.104.155431}{104}}(\href{https://link.aps.org/doi/10.1103/PhysRevB.104.155431}{15})
  \href{https://link.aps.org/doi/10.1103/PhysRevB.104.155431}{155431}

\bibitem{PhysRevA.103.L060201}
Rossini D, Ghermaoui A, Aguilera M~B, Vatr\'e R, Bouganne R, Beugnon J, Gerbier
  F and Mazza L 2021 {\em Phys. Rev. A\/} {\bf
  \href{https://link.aps.org/doi/10.1103/PhysRevA.103.L060201}{103}}(\href{https://link.aps.org/doi/10.1103/PhysRevA.103.L060201}{6})
  \href{https://link.aps.org/doi/10.1103/PhysRevA.103.L060201}{L060201}

\bibitem{10.21468/SciPostPhys.12.1.011}
Alba V 2022 {\em SciPost Phys.\/} {\bf
  \href{https://scipost.org/10.21468/SciPostPhys.12.1.011}{12}}(\href{https://scipost.org/10.21468/SciPostPhys.12.1.011}{1})
  \href{https://scipost.org/10.21468/SciPostPhys.12.1.011}{11}

\bibitem{PhysRevB.103.L020302}
Alba V and Carollo F 2021 {\em Phys. Rev. B\/} {\bf
  \href{https://link.aps.org/doi/10.1103/PhysRevB.103.L020302}{103}}(\href{https://link.aps.org/doi/10.1103/PhysRevB.103.L020302}{2})
  \href{https://link.aps.org/doi/10.1103/PhysRevB.103.L020302}{L020302}

\bibitem{Francesco_2021}
Tarantelli F and Vicari E 2021 Out-of-equilibrium quantum dynamics of fermionic
  gases in the presence of localized particle loss arXiv:
  \href{https://doi.org/10.48550/arXiv.2112.05180}{2112.05180}
  [cond-mat.quant-gas]

\bibitem{Moca_2021}
Moca C~P, Werner M~A, Örs Legeza, Prosen T, Kormos M and Zaránd G 2021
  Simulating lindbladian evolution with non-abelian symmetries: Ballistic front
  propagation in the \textit{SU}(2) hubbard model with a localized loss arXiv:
  \href{https://doi.org/10.48550/arXiv.2112.15342}{2112.15342}[cond-mat.str-el]

\bibitem{PhysRevB.102.100301}
Dolgirev P~E, Marino J, Sels D and Demler E 2020 {\em Phys. Rev. B\/} {\bf
  \href{https://link.aps.org/doi/10.1103/PhysRevB.102.100301}{102}}(\href{https://link.aps.org/doi/10.1103/PhysRevB.102.100301}{10})
  \href{https://link.aps.org/doi/10.1103/PhysRevB.102.100301}{100301}

\bibitem{fabio_quantum}
Franchini F 2017 {\em An Introduction to Integrable Techniques for
  One-Dimensional Quantum Systems\/} (Springer International Publishing)

\bibitem{gardiner2004quantum}
Gardiner C and Zoller P 2004 {\em Quantum noise: a handbook of Markovian and
  non-Markovian quantum stochastic methods with applications to quantum
  optics\/} (Springer Science \& Business Media)

\bibitem{doi:10.1080/00107510601101934}
Jacobs K and Steck D~A 2006 {\em Contemporary Physics\/} {\bf
  \href{https://doi.org/10.1080/00107510601101934}{47}}
  \href{https://doi.org/10.1080/00107510601101934}{279--303}

\bibitem{PhysRevB.105.064305}
Piccitto G, Russomanno A and Rossini D 2022 {\em Phys. Rev. B\/} {\bf
  \href{https://link.aps.org/doi/10.1103/PhysRevB.105.064305}{105}}(\href{https://link.aps.org/doi/10.1103/PhysRevB.105.064305}{6})
  \href{https://link.aps.org/doi/10.1103/PhysRevB.105.064305}{064305}

\bibitem{10.21468/SciPostPhys.7.2.024}
Cao X, Tilloy A and Luca A~D 2019 {\em SciPost Phys.\/} {\bf
  \href{https://scipost.org/10.21468/SciPostPhys.7.2.024}{7}}(\href{https://scipost.org/10.21468/SciPostPhys.7.2.024}{2})
  \href{https://scipost.org/10.21468/SciPostPhys.7.2.024}{24}

\bibitem{Glen_2020}
Mbeng G~B, Russomanno A and Santoro G~E 2020 The quantum ising chain for
  beginners arXiv:
  \href{https://arxiv.org/abs/2009.09208}{2009.09208}[quant-ph]

\bibitem{Calabrese_2012}
Calabrese P, Essler F~H~L and Fagotti M 2012 {\em Journal of Statistical
  Mechanics: Theory and Experiment\/} {\bf 2012}
  \href{https://doi.org/10.1088/1742--5468/2012/07/p07016}{P07016}

\bibitem{Heinz_2010}
Breuer H~P and Petruccione F 2010 {\em The Theory of Open Quantum Systems\/}
  (Oxford Scholarship Online)

\bibitem{PhysRevB.84.165117}
Rieger H and Igl\'oi F 2011 {\em Phys. Rev. B\/} {\bf
  \href{https://link.aps.org/doi/10.1103/PhysRevB.84.165117}{84}}(\href{https://link.aps.org/doi/10.1103/PhysRevB.84.165117}{16})
  \href{https://link.aps.org/doi/10.1103/PhysRevB.84.165117}{165117}

\bibitem{RevModPhys.80.517}
Amico L, Fazio R, Osterloh A and Vedral V 2008 {\em Rev. Mod. Phys.\/} {\bf
  \href{https://link.aps.org/doi/10.1103/RevModPhys.80.517}{80}}(\href{https://link.aps.org/doi/10.1103/RevModPhys.80.517}{2})
  \href{https://link.aps.org/doi/10.1103/RevModPhys.80.517}{517--576}

\bibitem{PhysRevB.94.245131}
Nandy S, Sen A, Das A and Dhar A 2016 {\em Phys. Rev. B\/} {\bf
  \href{https://link.aps.org/doi/10.1103/PhysRevB.94.245131}{94}}(\href{https://link.aps.org/doi/10.1103/PhysRevB.94.245131}{24})
  \href{https://link.aps.org/doi/10.1103/PhysRevB.94.245131}{245131}

\bibitem{lorenzo_quantum_2017}
Lorenzo S, Marino J, Plastina F, Palma G~M and Apollaro T~J~G 2017 {\em
  Scientific Reports\/} {\bf
  \href{https://doi.org/10.1038/s41598-017-06025-1}{7}}
  \href{https://doi.org/10.1038/s41598--017--06025--1}{5672} ISSN 2045-2322

\bibitem{calabrese2016quantum}
Calabrese P and Cardy J 2016 {\em Journal of Statistical Mechanics: Theory and
  Experiment\/} {\bf 2016}
  \href{https://doi.org/10.1088/1742--5468/2016/06/064003}{064003}

\bibitem{Alba_2022}
Alba V and Carollo F 2022 {\em Journal of Physics A: Mathematical and
  Theoretical\/} {\bf 55} 074002
  \urlprefix\url{https://doi.org/10.1088/1751-8121/ac48ec}

\bibitem{Alba_2022_1}
Alba V and Carollo F 2022 Logarithmic negativity in out-of-equilibrium open
  free-fermion chains: An exactly solvable case
  arXiv:\href{https://arxiv.org/abs/2205.02139}{2205.02139}

\bibitem{PhysRevB.105.144305}
Carollo F and Alba V 2022 {\em Phys. Rev. B\/} {\bf 105}(14) 144305
  \urlprefix\url{https://link.aps.org/doi/10.1103/PhysRevB.105.144305}

\end{thebibliography}

\end{document}